\documentclass[aps,pra,twocolumn]{revtex4}
\usepackage{psfig}
\newtheorem{proposition}{Proposition}

\def\trace{{\rm Tr}}
 
\def\TYP{TYP}
\def\more{\succ}
\def\hcal{{\cal H}}

\def\duzomniejsze{<\kern-.7mm<}
\def\duzowieksze{>\kern-.7mm>}

\def\textbf#1{{\bf #1}}
\def\beq{\begin{equation}}
\def\eeq{\end{equation}}
\def\be{\begin{equation}}
\def\ee{\end{equation}}
\def\ben{\begin{eqnarray}}
\def\een{\end{eqnarray}}
\def\beqa{\begin{eqnarray}}
\def\eeqa{\end{eqnarray}}
\def\eea{\end{array}}
\def\bea{\begin{array}}
\newcommand{\bei}{\begin{itemize}}
\newcommand{\eei}{\end{itemize}}
\newcommand{\bee}{\begin{enumerate}}
\newcommand{\eee}{\end{enumerate}}
\def\ra{\rangle}
\def\la{\langle}
\def\n{{\otimes n}}

\def\<{\langle}
\def\>{\rangle}

\def\dt#1{{{\kern -.0mm\rm d}}#1\,}
\def\ypodpis{\raise4mm\hbox{$\omega$}}

\def\rhon{\varrho^{\otimes n}}

\def\sigmam{\sigma^{\otimes m}}

\def\sigmamn{\sigma^{\otimes m_n}}

\def \s {\,\,\,\,}
\def\rab{\rho_{AB}}

\def\vr{\varrho}

\begin{document}
\title{Reversible transformations from pure to mixed states, and the unique measure of information}
\author{Micha\l{} Horodecki$^{(1)}$, Pawe\l{} Horodecki$^{(1)}$,
Jonathan Oppenheim$^{(1)(2)}$ }

\affiliation{$^{(1)}$Institute of Theoretical Physics and Astrophysics,
University of Gda\'nsk, Poland}
\affiliation{$^{(2)}$
Racah Institute of Theoretical Physics, Hebrew University of Jerusalem, Givat Ram, Jerusalem 91904, Israel}


\begin{abstract}
Transformations from pure to mixed states are usually associated with information loss and irreversibility.  Here,
a protocol is demonstrated allowing one to make these transformations reversible.  The pure states are diluted
with a random noise source.  Using this protocol one can study optimal transformations between states, and from this derive the
unique measure of information.  This is compared to irreversible transformations where one does not have access to noise.
The ideas presented here shed some light on attempts to understand entanglement manipulations and the inevitable irreversibility
encountered there where one finds that mixed states can contain "bound entanglement".
\end{abstract}
\maketitle

\section{Introduction}

There are two opposing pictures of {\it information}.
In the first picture, a source produces a large amount of information if
it has large entropy. Thus information can be associated with entropy.
This is because the receiver is being informed only if he is
``surprised''. In such an approach the information has a subjective meaning:
something which is known by the sender, but is not known by receiver.
The receiver treats the message as the  information,
if she didn't know it.

One can consider a different approach to information -- an objective one where 
a system represents information if it is in pure
state (zero entropy).  The state is itself the information.  This view is more natural in the context of
thermodynamics.  There, "knowledge is power" in the sense that 
one can draw work from a single heat bath
by use of systems in known pure states \cite{szilard}. On the other hand, the heat bath is represented by
a maximally entropic state, hence it is the less informative one. The pure
state represents information needed to order the energy of the heat bath.


There can be many candidates for functions
to measure information. However, Shannon recognized that
there is a unique function that shares some natural properties
to describe information.  Shannon, derived his unique measure
based on the subjective picture of information.
Therefore his  information  function (Shannon entropy) increases
as the dispersion of the probability distribution  increases.
The same is true of the generalization of Shannon's entropy to the quantum case
which  is the von Neumann's  entropy $S(\varrho)=-\trace \varrho\log\varrho$.

One can consider a measure of objective information, that has
the converse tendency: namely $I= \log d - H$ where $\log d$
is the maximal entropy of the  system (i.e.  the system has $d$ states).
In the quantum case it would be  $\log d - S(\varrho)$.
Such a function was naturally interpreted as the
information contents of the state as introduced by
Brillouin. 

One can ask the  question: can this function be
derived independently of the notion of  entropy, so that it is
not just a subtraction  of two known terms, but rather
has an autonomous meaning?

It turns out that there {\it is} such a possibility and it is offered by
quantum  information theory: in Ref. \cite{nlocc} we have
derived the function $I$ as the unique one, that does not increase
under some class of operations.  The motivation came from considering
information as a resource in distributed systems\cite{OHHH2001}.
The main aim of the present paper is to present the full rigorous version
of that derivation.  In the process, we give a protocol for reversible transformations
between states using a random source of noise. We also  discuss these
results in the context of the issue of reversibility and
entanglement theory.

It is quantum information theory (QIT) that
provides us with a suitable perspective to attack the problem.
Indeed, one of the central themes of QIT is the idea
of {\it optimal transitions between states under a restricted class
of operations}. This originates from attempts to describe
entanglement of quantum states. 
Although it was difficult to say what exactly entanglement was,
it was clear that it could not increase under the class
of operations made up
of {\it local operations and classical
communication} (LOCC)\cite{BBPSSW1996,BDSW1996}.  These operations allow one to use
any amount of separable states for free, but do not allow one
to create entangled states. One can take the converse  point of view:
one starts with a given class of operations (LOCC operations),
and treat the states that are not
free as containing  a resource, which can be called entanglement
(cf. \cite{Vidal-mon2000}).  The basic questions of entanglement theory
is: can state $\varrho$  be transformed into $\sigma$
by LOCC? What is the optimal rate
of such a transition?

In entanglement theory, this allowed one to define a number of measures
of entanglement, since essentially, any function which does not increase under LOCC
is a measure.  However, thus far, no one has found a unique measure.  The essential difficulty
(as will become clearer) is that operations under LOCC are not reversible.  However, if one has
a restricted class of operations for which transitions are reversible, then we will see that
the rate of transitions gives one a unique measure.  This is similar to pure bipartite state entanglement
where we have reversibility, and there is unique measure
of entanglement (entropy of subsystem)\cite{popescu-rohrlich, BBPS1996}.

In the present work, we consider a restricted class of operations we shall call 
Noisy Operations (NO) \cite{Karol}
and use this to develop a unique measure for information.  Essentially, we consider operations where one is
allowed to use random noise as a free resource.  Perhaps counter-intuitively, randomness allows one to make the
transformations reversible:  the number of pure states and noise
which is needed to form the state, is the same as the amount that can be obtained from the state.
The usual interpretation of mixed states, is that there creation involves
irreversibly destroying information.  Here we see that if one has access to noise as a resource,
then there is no irreversibility.

This has interesting
consequences concerning entanglement theory, since there, the
irreversibility is often associated with the fact that one is dealing with
mixed states.  Here, we see that transitions into mixed states need not
involve irreversibility.  In fact, the axiomatic structure of the paradigm
presented here involving mixed states is very similar to pure state
entanglement manipulation.  This shows that {\it apriori}, mixed state
entanglement manipulation need not involve irreversibility, leaving
open the question of why entanglement manipulation involves inevitable
irreversibilities.

Other restricted classes of operations may lead one to find unique measures for other quantities.
Here, we consider the optimal transitions
between states by means of NO. In the asymptotic limit of many identical
copies,  we will obtain that there is only one function that
does not increase under NO.  We will  establish that the optimal ratio
of conversion between a state $\varrho$ of a $N$ qubit system
and a state $\sigma$  of a $N'$ qubit system is equal to
$N-S(\varrho)\over N'-S(\sigma)$.   The transitions are reversible,
even though mixed states are involved.
Finally we will consider operations without free noisy ancillas.
Then the  mixed states have to be created from pure states
by partial trace, which introduces irreversibility.
We discuss the implications of our results on understanding
entanglement transformations, especially bound entanglement.

The work is organized as follows:
In Section \ref{sec:NO} we introduce the class of Noisy Operations.
Then in Section \ref{sec-single-copy} we show how one can transform a given state into
another state, under NO, provided certain conditions are met.  In \ref{sec:asymptotic}
we go to the asymptotic regime, and show that these transition rates are optimal.  This
will allow us to find the unique measure of information in Section \ref{sec:measure}.
In Section \ref{sec:revandirrev} we discuss the case of transitions without access to
noise, and give the transition rates in this case.  We discuss this in terms of understanding 
the source of irreversibility in transitions, and relate it to attempts to understand 
entanglement in Section \ref{sec:discussion}.  We
conclude with some open questions in Section \ref{sec:conclusion}.

\section{Noisy operations}
\label{sec:NO}

Perhaps the most important restricted class of operations which
has been considered in quantum information theory is  LOCC,
which was introduced
in the context of understanding entanglement in shared systems.
One is then interested in such questions such as how many maximally
entangled states
can a particular state be transformed into (i.e. the {\it rate}
of distilling singlet).
However, analyzing LOCC operations proved rather difficult.
Therefore,
to facilitate the investigation of  entanglement, a
larger classes of operations were analyzed
 -- so called
PPT operations\cite{Rains1999,Rains-erratum1999,PlenioVedral1998} which are superoperators
which preserve the positivity of partial transpose.

One can also consider other restricted classes of operations, and consider
various versions of the state transformation problem.
On the extreme end, one allows all operations, and adding any ancilla.
Then any state can be created for free, so that there is no resources
to be manipulated, and the theory becomes trivial.

As one knows  any operation can be composed out of a unitary operation,
adding an ancilla in some state, and removing ancilla.
Suppose that we want to make the theory nontrivial, while
keeping all unitaries in our class of allowable operations.
The only way is then to restrict the state of the free ancilla,
or somehow restrict removing ancillas.
In the present work, we consider only
restrictions to the free ancilla.  While one could instead
consider restrictions on removing ancillas, we believe that
this would give identical results\cite{context}.

Thus we will restrict to choosing states that can be
added for free by means of ancillas.  Remarkably, the choice of
which ancillas to allow is forced on us.
It turns out that the only choice that does not make the theory trivial
is that the free ancilla must be in maximally mixed state.  Essentially, we will
see in Section \ref{sec:discussion} that
if one allows any other ancilla, then all transition rates become infinite.
This fixes the class of operations we will call
{\it noisy operations} (NO).  The class NO is therefore very natural, as it
is the only one which gives non-trivial transition rates.


In entanglement theory, an entangled state of
Schmidt rank 2 represents the same
resource whether it acts on a Hilbert space $C^2\otimes C^2$
or on a larger space $C^d\otimes C^d$. This is because
embedding a state into a larger Hilbert space
is equivalent to adding local ancillas in a pure state.
In our case, a state acting on a Hilbert space $C^2$
is not the same resource as the one acting on
$C^d$. This is because adding ancilla in pure state
is adding a new resource.


\section{Optimal transitions under noisy operations - single copy case}
\label{sec-single-copy}

In this section we will present a protocol to transform single 
copies of states
into each other by diluting them with noise.  
We will show that the  transition  from  a single copy of
$\varrho$  to a single copy of state $\sigma$ is possible
if and only if the latter is {\it more mixed} than the former.
This is provided the Hilbert space is the same for both states,
i.e. they occupy the same number of qubits. We will also consider
the transitions between systems of different number of qubits.
One then has to add maximally mixed ancillas to one of the systems
(or to both), so that the number of qubits  become equal. Then
we can apply the above criterion.
The term ``more mixed'' \cite{Uhlmann-ordering} has the
following meaning:
For states $\varrho$ and $\varrho'$ on the Hilbert space $\hcal=C^d$, we say
that $\varrho$ is more mixed than $\varrho'$ ($\varrho\more\varrho'$) if
their eigenvalues in decreasing order satisfy
$\sum_{i=1}^k \lambda_k\leq
\sum_{i=1}^k \lambda'_k$, for all $k\leq \dim\hcal$.
(In the same way,   one can say that some probability distribution
is more mixed than another one).
If  the state is more mixed, its eigendistribution  is more spread.
 The order introduced by
the relation ``$\more$'' has a largest element -- the maximally mixed state.
It is easy to see that it is more mixed than any other state.

Let us now prove the main result of this section.

\begin{proposition}
\label{prop-more-mixed}
For states $\varrho$ and $\sigma$ of $d$-level  systems the
transition $\varrho \to \sigma$ by NO is possible
if and only if $\varrho \more \sigma$.
\end{proposition}

{\bf Proof.} ``$\Rightarrow$'' follows from  the fact
\cite{Chefles-major2002}
 that $\sigma\more \varrho$ iff
there exists a bistochastic map \cite{bistochastic} that
maps $\varrho$ into $\sigma$. Since noisy operations (for
equal input and output dimensions) are
bistochastic, then $\varrho\to\sigma$ implies $\sigma \more\varrho$.
To prove ``$\Leftarrow$'' we  cannot use the result of \cite{Chefles-major2002}, because
we do not know if the existing map can be taken to be noisy operations.
Instead we will construct the map explicitly. Let us then assume that
$\sigma\more\varrho$. First we can always rotate $\varrho$ unitarily, so
that it commutes with $\sigma$. Thus we can assume without loss
of generality   that the states commute. We can now use the fact
\cite{Bhatia} that if probability distribution
$\{q_i\}$ is more mixed than $\{p_i\}$, then the former can be
obtained from the latter via a mixture of permutations, i.e.
\be
q_i= \sum_j\alpha_j p_{\sigma_j(i)}
\ee
where $\sum_j\alpha_j=1$, while $\sigma_j$ are permutations of indices
of the probability distribution. Let  then $p_i$ be the eigenvalues of $\varrho$
and $q_i$ - eigenvalues of  $\sigma$. We will consider state
$\varrho\otimes\tau_N$  
(where $\tau_N$ is an added maximally mixed state of dimension $N$)
and  construct some permutation  of eigenvalues of the latter
density matrix.
After such permutation, and removing the ancilla, the state
will approach $\sigma$ for large $d$.
For simplicity we will assume that there are only two permutations
$\sigma_1$ and $\sigma_2$, so that $q_i=\alpha p_{\sigma_1(i)}+
(1-\alpha)p_{\sigma_2(i)}$.

The state $\varrho\otimes \tau_N$ consists of blocks, of dimensions $N$:
\be
\frac{1}{N}(\underbrace{p_1,\ldots,p_1}_N,\ldots,\underbrace{p_d,\ldots,p_d}_N)
\ee
We will divide each block into two groups of entries:
$N_1$ first entries and the rest $N_2=N-N_1$ entries.
Now we will apply permutation $\sigma_1$ to the
first entries of each block.  Similarly, we apply it to the second 
set of entries,
and so on, in the first group. The second group is subjected
to permutation $\sigma_2$ in a similar way. The resulting density
matrix is
\begin{widetext}
\be
\frac{1}{N}
(\underbrace{p_{\sigma_1(1)},\ldots,p_{\sigma_1(1)}}_{N_1},
\underbrace{p_{\sigma_2(1)},\ldots,p_{\sigma_2(1)}}_{N_2}
\ldots,
\underbrace{p_{\sigma_1(N)},\ldots,p_{\sigma_1(N)}}_{N_1},
\underbrace{p_{\sigma_2(N)},\ldots,p_{\sigma_2(N)}}_{N_2}
)
\ee
\end{widetext}
Now we trace out the ancilla. This means that we sum all elements
of each block, and instead of the block, take the resulting number.
The obtained eigendistribution   is given by
\be
\tilde q_i= {N_1\over N}  p_{\sigma_1(i)}+
{N_2\over N}p_{\sigma_2(i)}.
\ee
Choosing large $N$ and suitable $N_1$, $N_2$ one can
approach $\alpha$ and $1-\alpha$
with arbitrarily high accuracy. This ends the proof of
the proposition.

\section{Optimal transitions under noisy operations - asymptotic regime}
\label{sec:asymptotic}
Here we will consider asymptotic transitions 
of type 
\be
\rhon \to \sigmamn
\ee 
Usually it is not possible to obtain a perfect state $\sigmamn$ from
$\rhon$ even if an arbitrarily large amount of copies can be used.
It is however possible to obtain the state $\sigma_n$ that will
asymptotically converge to $\sigmamn$
\be
\rhon \to \sigma_n \approx \sigmamn
\ee
Thus we allow for inaccuracy, provided  it vanishes in
the limit of large $n$. The fidelity can be measured by the trace norm, i.e.
one requires that
\be
\|\sigma_n - \sigmamn\|\to 0\quad \mbox{for}\quad  n\to \infty.
\label{eq-asymp-accuracy}
\ee
The rate of given protocol of asymptotic $\varrho \to \sigma$
transition is given by the asymptotic ratio $\lim_n {m_n\over n}$.
The optimal transition rate denoted by $R(\varrho\to\sigma)$
is given by supremum over rates attainable by protocols that
satisfy the asymptotic accuracy condition (\ref{eq-asymp-accuracy}).

\subsection{Conversion from mixed to pure states}
We will now consider
the optimal rate for transition to the one qubit pure state $\pi$ i.e. $\varrho\to\pi$.
We will show that if $\varrho$  is a state of $d$-level system  then
\be
R(\varrho\to\pi)=I(\varrho)
\ee
where $I=N-S(\varrho)$ with
$N=\log d$ being the amount of qubits occupied by the state
$\varrho$.  In other words, the transformation from pure states to mixed states is reversible,
in the sense that the number of pure states which is needed, or which can be obtained
is the same.  The proof could be just use of Schumacher compression
\cite{Schumacher1995}, however with a different interpretation (similar
to that in \cite{VaziraniS1998}).
We will also  show that conversely, the amount of copies in state $\varrho$
that can be obtained under $NO$ per input pure qubit is also equal
to $I$.  The proofs will be similar to the reasoning
of Nielsen in \cite{Nielsen-pure-entanglement} where he derived
asymptotic  rates of pure state entanglement manipulations
from single copies based on majorization.

We will use law of large numbers \cite{CoverThomas,Schumacher1995}, that
implies that there exists a subset of eigenvalues of $\rhon$
call the typical set $\TYP$ with useful properties.
More precisely, given $\epsilon,\delta>0$, there exists large enough $n$,
and the set $\TYP$ of eigenvalues such that
\ben
&&\sum_{p_i\in\TYP} p_i\geq 1-\epsilon \\
\label{eq-typowe-eps}
&& 2^{-n(S+\delta)} \leq p_i\leq 2^{-n(S-\delta)} \quad \mbox{for} \quad
p_i\in \TYP
\label{eq-typowe-unif}
\een
These are thus the eigenvalues that carry almost the whole weight  and
they are more or less
uniform.  One can consider two states $\varrho_{typ}$
and $\varrho_{atyp}$, given by
\be
\varrho_{typ}={1\over c} \sum_{p_i\in\TYP} p_i|i\>\<i|,\quad \varrho_{atyp}
= {1\over 1-c} \sum_{p_i\not\in\TYP} p_i|i\>\<i|
\ee
where $|i\>$ are eigenvectors corresponding to $p_i$,
and $c=\sum_{p_i\in\TYP} p_i$ is a normalization constant.
Clearly $\rhon$ is a mixture of those states
\be
\rhon =c \varrho_{typ} + (1-c) \varrho_{atyp}
\ee
Since $c\geq 1-\epsilon$ one finds that $\varrho_{typ}$ is close to $\rhon$:
\be
\|\varrho_{typ}-\rhon\|\leq 2\epsilon
\ee
Thus it suffices to use $\varrho_{typ}$ instead of $\rhon$. Let
us first show that   one can convert $\varrho_{typ}$ into
approximately $n(N-S)$ copies of pure qubits. To this end, note that
the eigenvalues of $\varrho_{typ}$ satisfy:
\be
\lambda_i\equiv {p_i\over c} \geq {1\over c} 2^{-n(S+\delta)}.
\ee
Thus $\varrho_{typ}$ is less mixed than the state $\varrho_{out}$
with eigenvalues
\be
\{\underbrace{{1\over D}, \ldots, {1\over D}}_D,
\underbrace{0, \ldots, 0}_{d^n-D}\},
\ee
where $D$ is given by
\be
D=\left \lceil {1\over {1\over c} 2^{-n(S+\delta)}} \right \rceil
\label{eq-d}
\ee
(the eigenvectors of $\varrho_{out}$ are irrelevant, as we can perform
any unitary transformation for free).
Both of the states act on a $d^n$ dimensional space, so that we can apply our
Prop. \ref{prop-more-mixed}. Thus it is possible to go
from $\varrho_{typ}$ to $\varrho_{out}$ via noisy operations.
If we  choose $D$ to be larger than  in eq. (\ref{eq-d}), namely
so that it  is a power of $2$, the transition is still possible.
The smallest such $D$ satisfies
$\log D= \lceil n(S+\delta)\rceil\leq n(S+\delta) +1$.
Then the state $\varrho_{out}$ represents exactly the tensor product of
$\log D$ qubits  in maximally mixed state and
$n \log d -\log D\geq n(\log d - S - \delta) -1$
qubits in pure states. Thus one can remove the mixed qubits, and
keep the obtained pure qubits. Call the obtained state $\pi_{out}$.
 The rate of the transition
is the number of obtained pure qubits divided by $n$.
For large $n$ this tends to $\log d - S -\delta$. Since
$\delta$ can be chosen arbitrarily small, we obtain the optimal
asymptotic rate equal to $\log d - S$.

One could think that we obtain the pure qubits exactly. However,
we used Proposition \ref{prop-more-mixed}, where
the transition is not exact, though arbitrarily precise.

Yet we have not transformed $\rhon$ but $\varrho_{typ}$. We now take instead
of $\varrho_{typ}$, the state $\rhon$ and apply the same
action, which transformed $\varrho_{typ}$ into the required amount
of pure qubits (call the action $\Lambda$). It is now easy to see
that $\Lambda(\rhon)$ is close to a final state of pure qubits $\pi_{out}$.
Indeed, we have
\be
\|\Lambda(\rhon)-\pi_{out}\|=\|\Lambda(\rhon)-\Lambda(\varrho_{typ})\|
\leq \|\rhon - \varrho_{typ}\|\leq \epsilon
\ee
where the second last inequality comes from the fact
that completely positive trace-preserving maps
are contractions on Hermitian operators in trace norms, i.e.
$\|\Lambda(A)\|\leq \|A\|$ for Hermitian $A$ \cite{Ruskai-contr}.

Now we should show that the converse is possible, i.e.
to create a state $\rhon$ it is sufficient to start with
$\log d -S$ pure qubits per output copy of $\varrho$.
However, the proof is similar to the above. The only difference
is that we now use the other part of eq. (\ref{eq-typowe-unif}).
Namely, we note that $\varrho_{typ}$ is {\it more} mixed than
the state  with eigenvalues
\be
\{\underbrace{{1\over D'}, \ldots, {1\over D'}}_{D'},
\underbrace{0, \ldots, 0}_{d^n-D'}\},
\ee
where $D'$ is given by
\be
D'=\left \lfloor {1\over {1\over c} 2^{-n(S-\delta)}} \right \rfloor
\label{eq-d'}
\ee
Again, due to Proposition
\ref{prop-more-mixed} we can turn $\varrho_{typ}$ into the latter state.
Changing  $D'$ into a suitable power of $2$  (so that it is smaller
than  $D'$ of the above equation  hence passing from $\varrho_{typ}$
is still possible)   one gets that the latter
state is a tensor product of
$\log D'$ qubits in maximally mixed states and
approximately  $n(\log d -S)$ qubits in pure states.

Thus starting with $n(\log d -S)$ qubits in a pure state, one
has  to add $\log D'$ qubits in the maximally mixed state,  and
pass to the state $\varrho_{typ}$ which can be made arbitrarily close to
$\rhon$ by choosing small $\epsilon$.

\subsection{Optimality of $\log d - S$ transition rates and
optimal mixed-mixed transition rates }
We will now
show that the obtained rates are optimal. We will follow   Ref.
\cite{popescu-rohrlich} invoking standard thermodynamical
reasoning concerning Carnot efficiency (cf. \cite{BBPS1996}). Essentially, we will
show that $I=N-S$ cannot increase under NO maps, and then show that if our transitions
are not optimal, one could increase $I$ under NO.  We will
use the reversibility of our protocol, and also the asymptotic continuity property of von Neumann entropy.

We will prove optimality by contradiction. Suppose that
for the transition to pure qubits $\varrho\to \pi$ 
one can obtain a better rate than $R(\varrho\to\pi)=N-S$ (where $N=\log d$,
$\varrho$ acts on $C^d$).
Then one can run the following transition
\be
\pi\to \varrho \to \pi,
\ee
and obtain a rate of such transition which is more than 1. In other
words, employing $n<m(N-S)$  pure qubits, according to the assumption,
one gets $m$ pairs in state $\varrho$.
Then one can apply the protocol of the previous section to the $m$ pairs
of $\varrho$, to obtain $m (N-S)$ pure  qubits.
Thus one would be able to {\it increase} the number of pure qubits
from $n$ to $m$. Repeating the procedure one can obtain an arbitrary
number of pure qubits.

Now, we have to show that this is impossible.
This follows from the fact that $N-S$ cannot increase under NO maps.
Indeed, unitary maps do not change the quantity. Partial
trace of one qubit decreases $N$ by 1, and can increase entropy
at most by 1. Finally, adding a system in maximally mixed state,
increases $N$ by 1, but also increases entropy by 1.
Now, for $m$ pure qubits, $N-S=m$, while for $n$ qubits we have
$N-S=n<m$, thus the function $N-S$ must increase.

This is yet not the full proof, as we have made an implicit assumption,
that the final qubits are exactly pure states. In fact it is not true,
as all our conversions are only asymptotically true. However
the von Neumann entropy is asymptotically continuous, namely
for $N$ qubit states $\varrho$ and $\sigma$  we have
\cite{Fannes1973}
\be
|S(\varrho) - S(\sigma)|\leq N \|\varrho-\sigma\| + O(1)
\ee
In our case we take $\varrho=\pi^{\otimes m}$ and
$\sigma_m$ being the actual final state. We know then that $S(\varrho)=0$
and that $\|\sigma_m - \varrho\|$ tends to zero as $m$ goes to infinity.
Thus ${|S(\sigma_m)|\over m}\to 0$  for large $m$. Thus
the {\it density} of the function $I$ tends to $1$
for the state $\sigma_m$. This density is also $1$ for the
initial state $\pi^{\otimes n}$.  Thus we can write that
in our process $I_{out}= m_n-o(m_n)$; on the
other hand $I_{in}=n$. We will show that for large $n$, (which
also implies that $m_n$ is large) $I_{in}<I_{out}$.
Indeed that latter inequality is equivalent to the following
set of equivalent inequalities
\ben
&& m_n-o(m_n)> n\\ \nonumber
&& {m_n\over n}-{o(m_n)\over n}> 1\\ \nonumber
&& {m_n\over n}(1-{o(m_n)\over m_n})> 1\\ \nonumber
\een
The quantity inside the bracket tends to 1, while in our protocol
$m_n\over n$ goes to a number greater than one. Thus the inequality
holds, which is impossible. Therefore our assumption
that our rate is not optimal is incorrect. In a similar way one can show that
one cannot obtain a better rate than
\be
R(\pi\to\varrho)={1\over I}
\ee
while going from pure states to mixed ones.

Clearly since the transitions from mixed to pure states are reversible and optimal,
one can use these protocols to go from one mixed state to another in a reversible
and optimal way by just distilling pure states and then creating another mixed state.
This gives that the optimal ratio
of conversion between  state $\varrho$ of a $N$ qubit system
and state $\sigma$  of a $N'$ qubit system is equal to
\beq
R(\varrho\to\sigma)=N-S(\varrho)\over N'-S(\sigma)\s .
\eeq
\section{Information monotones and the unique measure of information}
\label{sec:measure}

Here we will derive the unique measure of information $I$, with virtually no
assumptions. The derivation will be mostly operational.
We will actually assume two properties. The first will concern
the intuition of what information is -- namely, noisy operations
should not increase it. Indeed, information, whatever it is,
shouldn't be increased by unitary operations, by adding a
qubit in maximally mixed state (supposed to be information-less)
and discarding qubit (rather obvious requirement). Thus we postulate

{\bf Postulate 1.}
$I$ should be monotonic under noisy operations.

We will actually see in the next section, that this postulate is
rigid, in the sense that if instead of noisy operations,
we had chosen operations with a free resource
other than maximally mixed states, the theory would be trivial,
and all rates would be infinite.

The second assumption will not be connected with the expected properties of
information. Rather it will display the properties any function
used in the asymptotic regime (limit of many copies) should possess.  I.e.

{\bf Postulate 2.}  $I$ is asymptotically continuous.

By asymptotically continuous, one means that for the state $\varrho_N$
and $\sigma_N$  of $N$ qubits, such that $\|\varrho_N-\sigma_N\|\to 0$
for $N\to \infty$. One would then require
\be
|f(\varrho_N)-f(\sigma_N)|\to 0.
\ee
We then say that $f$ is {\it asymptotically continuous}.
The motivation for this is that
in the asymptotic regime, one identifies the states that asymptotically converge
to each other. Thus the only relevant functions of states
are those that also somehow identify those states. Of course
in the asymptotic limit, the interesting functions become infinite, so that
one has to pass to intensive quantities and divide by the number of copies to obtain {\it densities}.
The relevant functions would be those whose densities converge
on convergent sequences.
Note that this not merely a technical
requirement. Rather this follows from the basic assumption of the asymptotic
regime - that similar states should be identified. The latter
assumption is necessary, and physically natural -
it is simply impossible to obtain exact transitions.

Let us now prove  that there is a unique function that satisfies
these two postulates.

The proof can be obtained from Refs.
\cite{DonaldHR2001,Michal2001}. According to \cite{Michal2001}
The following inequality is true
\be
R(\varrho\to\sigma)\leq {f^\infty(\varrho)\over f^\infty(\sigma)}
\ee
where $R$ denotes the rate of transition under any given class of
operations, and $f$ is an asymptotically continuous
function nonincreasing under the class.  The symbol $\infty$ stands for
{\it  regularization}. The regularization of function $f(\varrho)$  is
$M^\infty(\varrho)=\lim_{n\to \infty}{1\over n}M(\varrho^\n)$.

Choosing as $\sigma$ the one qubit pure state $\pi$ and exchanging the roles of
$\varrho$ and $\sigma$ we obtain
\ben
&&R(\varrho\to\pi)\leq {f^\infty(\varrho) \over f^\infty(\pi)}\\ \nonumber
&& R(\pi \to \varrho)\leq {f^\infty(\pi)\over f^\infty(\varrho)}\\ \nonumber
\een
Denoting $1/f^\infty(\pi)=a$  we obtain
\be
R(\varrho\to \pi) \leq  a f^\infty(\varrho) \leq {1\over R(\pi\to\varrho)}
\ee
However we have explicit protocols which show
that $R(\varrho\to \pi)\geq I$ and $1/R(\pi\to\varrho)\leq I$.
Thus up to the constant $a$ we obtain that $f^\infty=I$. In this sense $I$ is the
{\it unique} measure of information.

It is interesting to see how  other measures of information
are removed in the asymptotic limit. Suppose that we consider measures of  information
which only satisfying the first postulate.
Since we see that everything is very similar to the
problem of pure state entanglement, one is not surprised that
all monotones under NO are so called Shur concave functions
of the density matrix.
In particular there is a set of information measures (or "monotones")
which is enough to determine if a transition is possible.
These are the so called Ky Fan k-norms, i.e. sums of the first $k$
largest eigenvalues. By definition
of the ``more mixed'' condition, we have $\varrho\more\sigma$ iff
for all $k$-norms, $||\varrho||_k\leq ||\sigma||_k$. Thus
the process $\sigma\to \varrho$ is possible  iff in the process
no monotone increases.

One might get the feeling that there is some contradiction here.
Namely, in asymptotic transitions, the only restriction
for the rate is the monotone $I$. Thus there are allowed transitions
for which other monotones increase. Indeed, we say that
$\rhon\to\sigmam$ is possible, though it is clear that
some of the monotones will increase. The solution is
that, in fact we are not  talking about exact transitions.
Thus in the actual transition, the final state obeys
the nonincreasing of monotones. For that state, all monotones are not
greater than for the initial state. The monotones are however not
asymptotically continuous, and they see differences between
that actual state, and the required state $\sigmam$.
The only monotone that does not see the difference is $I$.
Therefore only this function survives in the asymptotic limit.

\subsection{The choice of free resource is unique}
One could think that the way we have obtained the information measure
is not fully operational, as we assumed, somewhat arbitrarily
that the free resource is the maximally mixed state. Here we will show
that this is the only  reasonable choice, if we want to
allow ancillas at all,  and if the theory is to be nontrivial
i.e. the transition rates are finite,
and therefore, not all states can be obtained for free. We thus assume that our operations
include unitary transformations, and partial trace, and will try to
play with third component - adding ancillas.

Suppose that instead of maximally mixed states $\tau$, we chose any other state $\varrho_0$
as a free resource.  This means that we can use arbitrarily
many copies of this state. From $\varrho_0^{\otimes n}$ we
can produce {\it without use of noise} pure states by Schumacher compression
\cite{Schumacher1995,VaziraniS1998} (in this paper we
have not described this - we always used noise).  Thus we have pure states
for free.  From pure states we can produce noise by entangling two qubits
in a maximally entangled state and rejecting one qubit. The remaining
one will be in a maximally mixed state. This is not very efficient: we
spend two qubits in a pure state to get one qubit of noise. However
pure states are for free, hence this method is sufficiently good
in our situation. Now we have both noise and pure states
for free, hence via the protocol described in the previous section,
we can create any state. The theory becomes trivial - all states are
for free; all rates are infinite. Thus if we allow
adding systems for free at all, we can only add ones in maximally
mixed state.
We thus see that Postulate 1 is rather rigid, in the sense that changing it to a class
of operations which allows any other ancilla, will result in a trivial theory.

\section{Reversibility and irreversibility}
\label{sec:revandirrev}
Note that we have a kind of reversibility: the amount of pure
qubits that can be drawn from a given state is equal to the amount we need
to create the state.
Let us consider another situation, where we count everything (no free resource).
we then see that there is basic irreversibility: transitions from
almost any state $\varrho$ to any other state is irreversible.
For example, one can draw $I$ pure qubits from $\varrho$,
but to create $\varrho$, one needs many more pure qubits. There
are two reasons for this. The first reason is trivial - to get
$N$ qubits in state $\varrho$  
one needs $N$ qubits  anyway.  This is
1 qubit per output qubits, which is already more than $I=N-S(\varrho)$.
Now, however, even more pure qubits are needed. Namely, the output state
has nonzero entropy. However the only way of producing entropy
out of pure states is rather wasteful: one entangles two qubits,
and removes one of them (as already described in the previous section).
Indeed, previously, we had a free source of entropy - maximally mixed states,
now we have only pure states to our disposal, and we count them.

Interestingly, in the classical world there is no way to produce entropy at all.
Therefore in classical statistical mechanics, one has
to assume mixed state from the very beginning.
Quantum mechanics allows one to produce mixed states out of pure ones.
This may lead one to prefer Bayes concept of probability.

We will now show that
\begin{proposition}
$N+S$ pure qubits are necessary and sufficient to produce $\varrho$ if one
doesn't have access to noise.
\end{proposition}

That this is sufficient can be seen by noting that $\varrho$ can
be creating by preparing the {\it purification}
of $\varrho_{typ}$. We thus consider a pure state
of two systems $A$ and $B$. Subsystem $A$ has $N$ qubits, and
its state is $\varrho_{typ}$. The state of subsystem $B$ (the purification) is also $\varrho_{typ}$, but we
do not need it to be an $N$ qubit system, but rather want it to
use the smallest possible amount of qubits. The latter is equal to $S$
qubits. Thus $N+S$ qubits in pure state are needed to prepare
$\varrho_{typ}$  (preparation is discarding the system $B$).
That this number of qubits are necessary simply stems from the fact that we start from
an initially pure state, so to get a mixed state we must trace
out part of the initial system, and the ``garbage'' that gets traced out
must have at least $S$ qubits (since the number of qubits of garbage
cannot be less than it's entropy, and the garbage
must have entropy $S$ since the system is initially pure).  We must
also have at least $N$ qubits left over to form the state.
So, in general, to create the N qubit state $\varrho$
we need $N+S$ pure qubits, but we can draw only $N-S$ qubits.
The ``information of preparation'' is much greater than ''information
of distillation''. During the transition
\be
\psi \to \varrho \to \psi
\ee
we lose $2S$ pure qubits.

\begin{proposition}
\label{mix-mix-nonoise}
To produce the  mixed-mixed transition $\vr\rightarrow\sigma$, without access to noise,
$\Delta N+\Delta S$
qubits are necessary and sufficient
where $\Delta N \equiv N(\sigma)-N(\vr)$ and $\Delta S \equiv S(\sigma)-S(\vr)$
\end{proposition}

To see these resources are necessary, we note that a general protocol involves an initial
state $\vr\otimes|\psi\ra\la\psi|$, where $|\psi\ra$ is some initial pure state.
One then performs unitaries to give a state $\varrho'$, and then
one traces out the garbage $g$ to leave the state $\sigma$.
We can then use the triangle inequality
\beq
|S(\sigma)-S(g)|\leq S(\varrho')=S(\vr)
\eeq
to see that the number of garbage bits traced out $N(g)$ satisfies
$N(g)\geq S(g)\geq \Delta S$ (if $S(g)\geq S(\sigma)$ then trivially $S(g) \geq \Delta S$.
So, we need a minimum of $N(\sigma)+\Delta S$ pure qubits to create $\sigma$,
but we already had $N(\rho)$ bits to start, so the minimum amount of additional qubits needed is
$\Delta N+\Delta S$.

The protocol which realizes this bound is to reversibly distill
$\varrho$ into $N(\varrho) -S(\varrho)$ pure qubits
and $S(\varrho)$ bits of noise in a manner which we shall shortly describe.
We then add in an additional $\Delta N-\Delta S$ pure qubits.
However, we also need $\Delta S$ bits of noise, which costs $2\Delta S$ pure states
(this is the only part of the protocol which is irreversible).  We then
create $\sigma$ reversibly as described in the previous section, using the $\Delta N + \Delta S$
additional qubits.

The distillation procedure can be realized using a scheme
similar to quantum data compression\cite{Schumacher1995}
and to the concentration of entanglement scheme of Ref.
\cite{BBPS1996} (here however, the procedure is
applied to the entire state).  The protocol is essentially
a projective measurement onto blocks proportional to the identity.
On average, the size of the Hilbert space
that the state is projected onto will be of size $S(\varrho)$, and so,
the state can then be unitarily rotated to leave $N(\varrho) -S(\varrho)$
pure states.  We will explicitly give the protocol for $n$ qubits i.e. $N(\varrho)=1$
but the extension to higher dimensional states is straight-forward.

We can write the state in the eigenbasis which we label as $0$ and $1$, i.e.
$\varrho = a|0\ra + b|1\ra $.  We have have $n$ copies, i.e. we operate on
the state $\varrho^{\otimes n}$,
and then we measure how
many zeros this state has.  This is a
measurement with $n+1$ outcomes and it will yield a result $\quad k = 0,...,n$ telling us how
many zeros there are.  This projects us onto a state which
has $ d_{k} = \left({n \atop k}\right)$ basis vectors, all with equal coefficients.  I.e.
it is proportional to the identity.
The probability of finding a particular outcome $k$ is
$ p_{k} = \left({n \atop k}\right)a^{2k}b^{2(n-k)} $
and since it does not in general span the entire Hilbert space,
can be unitarily transformed to yield $I_k=n-\log{d_k}$ pure states.

Each process $ \varrho \to \{ p_{k}, \rho_{k}\} $
after which $I_{k}$ pure states is extracted from $\rho_{k}$ with
probability $p_{k}$,
provides
\beq
N_o=\sum_{k}p_{k}I_{k} - H(\{p\})
\eeq
total pure states.
The Shannon entropy $H(\{p\})$ of distribution $\{p_{k}\}$
equals the cost of the erasure of information which allows
us to work with an ensemble of $\rho_{k} $'s \cite{landauer}.
Thus we need $ I_{er} = H(\{p\}) $
bits of erasure to pay for the next part of the scheme, in which they  draw
$\sum_kp_kI_{k}$ pure states.  This quantity, which is of order $\log n$ is
negligible in the large $n$ limit.  We can divide the above equation by $n$ to obtain
the amount of extractable pure states per qubit.
\beq
N_o/n = 1 - S(\varrho)
\ee
where the erasure cost has been neglected since it is of order $\log n/n$.
This completes our proof of the proposition.


This allows one to think of states in the following way: the mixed state
consists of $N-S$ bits of information and $S$ bits of noise.
Thus to produce it one needs $N-S$ qubits in pure states, to account for
information, and $2S$ qubits to produce noise. Indeed one bit of noise costs
two pure qubits - since noise is produced by rejecting part of entangled
system.




It is interesting that one needs to add a free resource (noise) in order
to achieve efficient transitions from pure to mixed states which are much less ``useful''
than mixed-to-pure transitions. Indeed, the latter is a task
that can be associated with such actions as cooling, error-correction,
increasing signal. This useful task can be performed {\it without} the help of
an additional resource at the optimal rate. Only the converse direction,
which is not useful (who wants to have mixed states instead of pure ones?)
needs noise, and is much less efficient without noise.

There are other cases where reversibility needs noise. For
example according to the Shannon second theorem, one can simulate
one use of noiseless channel
by $1/C$  uses of a noisy channel of capacity $C$. However, one cannot
do the converse, i.e. simulate noisy channels by noiseless
one, without sharing random correlated data \cite{Shor}. Again, the
useful task does not need any additional resource, while the useless
task needs one. This is clear, if one realizes that in both situation
we deal with dilution of some valuable resource into noise.
Similarly in thermodynamics, the the thermodynamical system
with difference of temperature can be thought as being
``pure energy'' (such as mechanical energy) diluted into
``pure heat''. To draw work out of it one does not need any additional
resource. However to create the system of heat baths efficiently, one needs a heat reservoir
at the beginning. Otherwise, one has to spend work to produce heat, exactly
as we needed to spend pure states to produce noise.

\section{Discussion: comparison with entanglement transformations}
\label{sec:discussion}

The paradigm discussed in this paper may be useful to understand
the problems of entanglement theory.
As one knows there is a basic irreversibility in entanglement
transformations. We deal there with bipartite systems, shared
by distant parties. One is interested in how many 
pure singlets are needed to form a state $\rab$ (the 
{\it entanglement cost}, and
also, how many singlets can be obtained from the state
(the {\it distillable entanglement}).
If $\rab$ is pure, then the entanglement cost is equal
to the distillable entanglement in the limit of many copies
of $\rab$ \cite{BBPS1996}.  The tranformations are reversible.   However
it is known that for a number of mixed states, the
distillable entanglement is not equal to the entanglement
cost \cite{Vidal-irrev2001}.  One has irreversibility.  It has
generally been assumed that this is because one is making
transformations between pure states (in this case, singlets),
and the mixed state $\rab$.  One therefore expects some 
information loss.  However, as we have seen here, one can
make transformations between pure and mixed states 
completely reversible, provided one has access to noise.
And indeed, in the paradigm of entanglement theory,
there is no reason why two distant parties couldn't
share some initial noisy resource.   There
is no special {\it apriori} reason for irreversibility in 
entanglement theory.  It is therefore
interesting to compare the situation discussed here,
with that of entanglement theory.  This comparison 
is summerized in the following table, and described below

\begin{widetext}
  
\begin{tabular}{|l||l|l|l|l|}
\hline
Paradigm & Class of Operations & Free Resource & Expensive Resource
& Reversible? \\
\hline  \hline
information & NO & maximally mixed states & pure states & yes \\
\hline
pure state entanglement & LOCC & separable states & singlets & yes \\
\hline
mixed state entanglement & LOCC & separable states & singlets & no \\
\hline
thermodynamics & adiabatic processes & heat\cite{heat}
 & work & yes  \\
\hline
ToE\cite{toe} \cite{thermo-ent2002}
& LOCC + PPT states & PPT states & singlets & no (?) \\
\hline
PPT \cite{APE} & PPT operations &  PPT states & singlets & in some cases\\
\hline
\end{tabular}

\vspace{.2cm}
\end{widetext}

Instead of NO, in entanglement theory we have there LOCC, which means
that 1) arbitrary {\it local} unitary  operations can be performed,
2) any {\it local} ancilla  can be added, 3) any {\it local} partial
trace can be performed 4) qubits can be communicated between
distant parties only via a {\it dephasing channel}.
The role of noise is played by separable states - all the states
that can be produced for free within the allowed class of operations
are a free resources. The role of pure states is played by pure
entangled states.

One could imagine that like with "local information theory",
in entanglement theory, any state is a
reversible mixture of two phases: pure entanglement and a separable 
noisy phase.
One should be able to draw the same amount of pure entanglement
from a given state as is needed to produce it.  Creation of mixed states
would be reversible {\it dilution} of pure entanglement
into mixed, separable states.

In this simple picture we would have only two kinds of basic elements
in entanglement theory: pure entangled systems and  disentangled systems.
One is useful, the other - useless.
A state which is neither pure entangled nor disentangled,
consists of those two basic elements. This is in parallel to
the paradigm presented in this paper, where the useful elements
were pure states, the useless maximally mixed ones.



As noted, such a situation exists for pure states,
where we can reversibly concentrate and dilute entanglement.
However, such a situation does not exist with mixed states in entanglement
theory.  What is the basic difference between mixed state entanglement
and  the paradigms (I) of pure entanglement, and (II) the present NO one?

In both I and II we have the following common point. We define
states that can be added for free, and then the class of operations.
Then in both cases it turns out that the free states remain the only
nontrivial set of states closed under the class of operations.
Now in mixed state entanglement we may have another basic element -
bound entangled ones. One cannot obtain them from
separable states, but also one cannot obtain any pure entanglement
from them. Thus the set of  states closed under the class
of operations is greater than it would
seem from the construction of the paradigm.
Thus in situations I and II we have only two elements: useful
and useless. In paradigm II  the useful element is information, 
the useless one -
noise. In paradigm II - the useful element is entanglement, the useless -
separability. Here, entanglement {\it itself} is divided into 
at least to phases:
bound and pure. From bound entanglement we cannot make pure, so call it
useless as well. Thus we can have states that have entanglement,
but are useless. This is different than I and II, but similar to
thermodynamics: we have there two forms of energy, useful and useless.
In Ref.  \cite{thermo-ent2002} we have asked a question - is it possible
that mixed-state entanglement is like thermodynamics. There
would be three basic elements: separable states (no entanglement),
 bound entanglement and pure entanglement, similarly as
in thermodynamics there are states without energy, with disordered
energy (single heat bath) and with ordered energy (mechanical energy).
All three kinds could be reversibly mixed.

In \cite{thermo-ent2002} it was shown that such a picture can be treated 
as a sort of ``first order approximation'' rather than 
full description of asymptotic
bipartite entanglement.
Related questions were studied in Ref. \cite{APE} 
where reversibility for some states holds, if so
called PPT superoparators
are allowed \cite{Rains1999}. The relation between the latter result
and the ``thermodynamic'' approach of Ref. \cite{thermo-ent2002} 
goes beyond the scope of this paper and is explained in 
\cite{thermo-ent2002} itself.





\section{Conclusion}
\label{sec:conclusion}
Contrary to what might be imagined, we have shown that
mixed states do not necessarily impose irreversibility.  One can reversibly transform
pure states into mixed ones, provided one has access to random noise.  This defines
a class of operations (NO) which can then be used to explore the transition rates
between various states.  It is found that the information measure $I=N-S$ cannot
decrease under NO, and is therefore the unique asymptotically continuous measure
of information.  It would be extremely interesting to explore other restricted classes of
operations in addition to $NO$, to see whether there are other non-trivial theories.
Exploring the connection between this, and the LOCC paradigm of entanglement theory,
would be extremely useful in understanding entanglement in distributed quantum systems.
Perhaps ideas along the lines of \cite{thermo-ent2002} may prove 
fruitful.

\begin{acknowledgments}
MH and PH thank  Ryszard Horodecki and JO thanks Jacob
Bekenstein for numerous discussions
on the notion of information.  
This work is supported by EU grant EQUIP, Contract No. IST-1999-11053.
JO acknowledges the support of
the Lady Davis Fellowship Trust, and
grant No. 129/00-1 of the Israel Science Foundation.
\end{acknowledgments}

\begin{thebibliography}{32}
\bibitem[{con()}]{context}
\bibinfo{note}{See in this context, \cite{OHHH2001} where we introduced Closed
  Local Operations and considered the case where one is not allowed to remove
  ancillas}.
\bibitem[{bis()}]{bistochastic}
\bibinfo{note}{A linear completely positive map is called bistochastic if it
  preserves both trace and identity.}
\bibitem[{hea()}]{heat}
\bibinfo{note}{Here, heat is understood as a single heat bath; two baths at
  difference temperatures are not "pure" heat in the sense that they can be
  used to perform work.}
\end{thebibliography}

\end{document}